\pdfoutput=1

\documentclass[11pt]{article}

\usepackage[final]{acl}

\usepackage{times}
\usepackage{latexsym}
\usepackage{multirow}

\usepackage[T1]{fontenc}

\usepackage[utf8]{inputenc}
\usepackage{graphicx}
\usepackage{subfig}
\usepackage{appendix}
\usepackage{microtype}

\usepackage{inconsolata}
\usepackage{booktabs}
\usepackage{enumitem}
\usepackage{amsmath}

\newcommand*\samethanks[1][\value{footnote}]{\footnotemark[#1]}
\newtheorem{myDef}{Definition}
\title{Make Large Language Model a Better Ranker}

\author {
    Wen-Shuo Chao\textsuperscript{\rm 1},
    Zhi Zheng\textsuperscript{\rm 2},
    Hengshu Zhu\textsuperscript{\rm 1,\rm 3}\thanks{Corresponding Author},
    Hao Liu\textsuperscript{\rm 1}\samethanks \\
    \textsuperscript{\rm 1} The Hong Kong University of Science and Technology (Guangzhou) \\
    \textsuperscript{\rm 2} School of Data Science, University of
 Science and Technology of China \\
    \textsuperscript{\rm 3}  Computer Network Information Center, Chinese
 Academy of Sciences \\
    \{wschao829, liuh\}@connect.hkust-gz.edu.cn, \\
    zhengzhi97@mail.ustc.edu.cn,
    zhuhengshu@gmail.com
}

\begin{document}
\maketitle
\begin{abstract}

Large Language Models (LLMs) demonstrate robust capabilities across various fields, leading to a paradigm shift in LLM-enhanced Recommender System (RS).
Research to date focuses on point-wise and pair-wise recommendation paradigms, which are inefficient for LLM-based recommenders due to high computational costs. 
However, existing list-wise approaches also fall short in ranking tasks due to misalignment between ranking objectives and next-token prediction.
Moreover, these LLM-based methods struggle to effectively address the order relation among candidates, particularly given the scale of ratings.
To address these challenges, this paper introduces the large language model framework with Aligned Listwise Ranking Objectives (ALRO). ALRO is designed to bridge the gap between the capabilities of LLMs and the nuanced requirements of ranking tasks.
Specifically, ALRO employs explicit feedback in a listwise manner by introducing soft lambda loss, a customized adaptation of lambda loss designed for optimizing order relations.
This mechanism provides more accurate optimization goals, enhancing the ranking process.
Additionally, ALRO incorporates a permutation-sensitive learning mechanism that addresses position bias, a prevalent issue in generative models, without imposing additional computational burdens during inference.
Our evaluative studies reveal that ALRO outperforms both existing embedding-based recommendation methods and LLM-based recommendation baselines.

\end{abstract}

\section{Introduction}

The rapid advancement in Large Language Models (LLMs), known by GPT-4~\cite{achiam2023gpt}, has marked a significant milestone in demonstrating their versatility in zero-shot and few-shot learning across various domains. 
These models, effectively employed in domains like Question Answering and Information Retrieval, have shown remarkable adaptability and reliability. 
Their ability to efficiently handle tasks usually requiring extensive domain-specific training has sparked a surge in research aimed at exploring their potential across diverse applications, e.g. Recommender System.


\begin{figure}[t]
    \centering
    \includegraphics[width=0.48\textwidth]{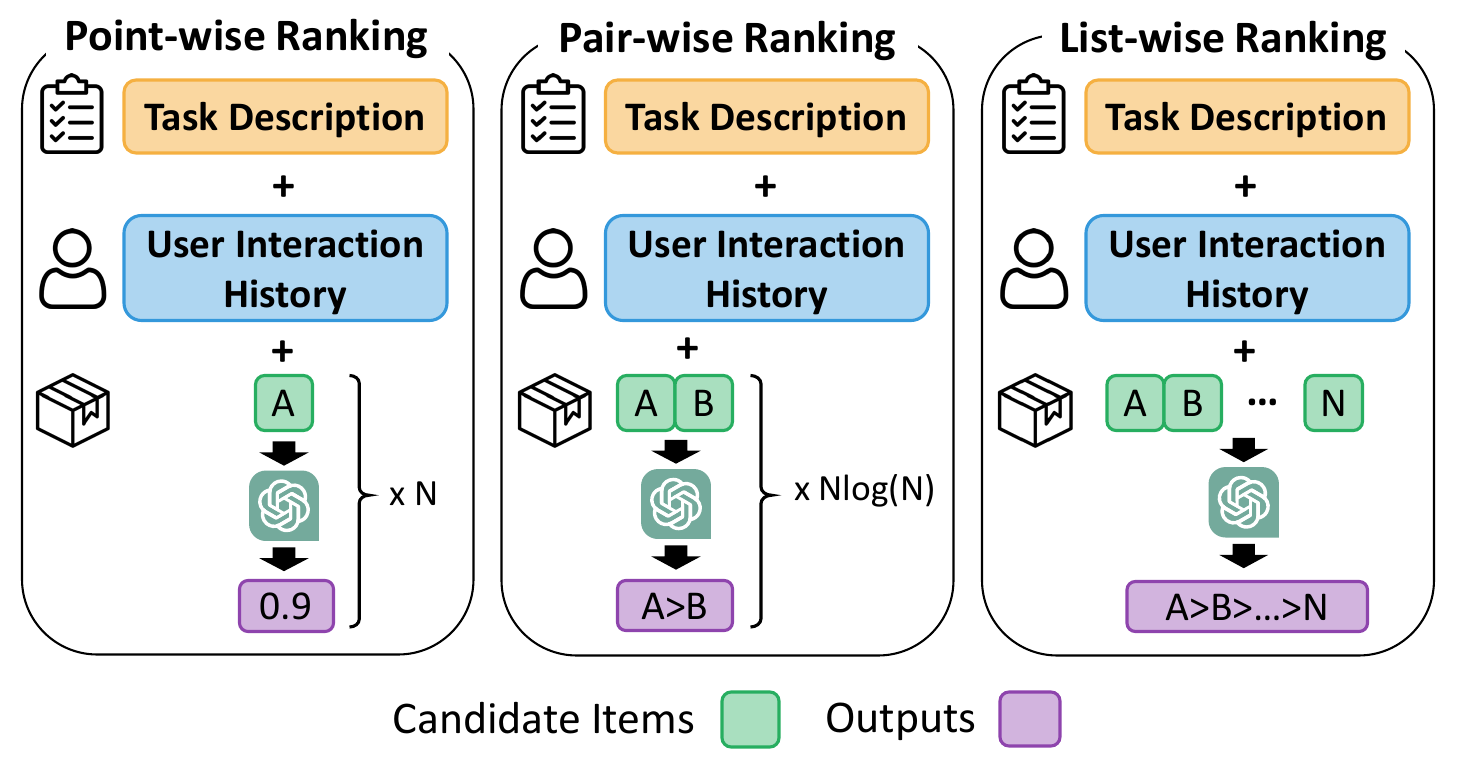}
    \caption{The comparison of point-wise, pair-wise, and list-wise ranking in LLM-based recommendation.}
    \label{fig:comparison}
\end{figure}

In the context of recommender systems, the application of LLMs has attracted considerable attention. 
\citet{wu2023survey} demonstrates a novel paradigm in using Large Language Models as recommender systems. This approach leverages the natural language processing strengths for context-sensitive recommendations. 
Concurrently, investigations conducted in \citet{bao2023tallrec} and \citet{li2023e4srec} explore the capability of LLM in point-wise recommendation, revealing how language models can be adapted for suggesting products. 
\citet{qin2023large} investigate pairwise ranking prompts to enhance recommendation systems.
Despite these advancements, as depicted in Figure~\ref{fig:comparison}, a significant limitation of these methods is their high computational cost, stemming from the iterative call of LLMs to evaluate each candidate item.
Moreover, existing approaches focus on implicit feedback, filtering rating signals with predefined thresholds. 
This practice fails to effectively address partial order relations inherent in the magnitude of ratings.

In leveraging LLMs for recommendation systems, the list-wise ranking method stands out for its computational efficiency~\cite{yue2023llamarec,yang2023palr}. 
However, executing list-wise ranking with explicit feedback effectively is fraught with challenges~\cite{dai2023uncovering}.
The core issue lies in the objective misalignment between LLMs' natural language generation and ranking tasks. 
Specifically, ranking demands a sophisticated reasoning process to understand partial order relation within the sequence of candidates based on the ratings, which cannot be addressed by supervised fine-tuning with cross-entropy~\cite{dai2023uncovering, xu2024fairly}.
Optimizing Large Language Models to interpret the magnitude of these ratings and the order relation among candidates remains a critical challenge in enhancing the ranking performance.
Additionally, the inherent position bias in LLM-generated lists further complicates the matter. 
This bias indicates that the initial input ordering of the candidates significantly influences the final ranking of potential outputs.
Although techniques like bootstrapping, suggested by~\citet{hou2023large}, offer a solution by iteratively querying the LLM with permuted candidate sequences to obtain unbiased arrangements, this method significantly increases computational demands. 
Such an increase is particularly problematic given the substantial resources required by Large Language Model operation, thereby highlighting a crucial trade-off between the precision and practicality of employing LLMs as recommender systems.





To overcome the aforementioned challenges, we propose the Large Language Model learning Framework with Aligned Listwise Ranking Objectives (ALRO), which integrates explicit feedback and soft lambda loss and permutation-sensitive learning into the training process to enhance the ranking capabilities of Large Language Models (LLMs). 
This enhancement is achieved through supervised fine-tuning and Low-Rank Adaptation (LoRA)~\cite{hu2021lora}.
Specifically, ALRO employs a soft lambda loss that effectively bridges the gap between the objectives of ranking and language generation.
This transformation emphasizes the significance of item orders within the predicted list, augmenting their impact during the language generation task.
Furthermore, we introduce a permutation-sensitive learning framework designed to enhance ranking consistency by evaluating the distance between outputs from permuted candidate lists, thereby ensuring stable ranking outcomes regardless of candidates' input order.
This strategy boosts the permutation invariance capability of the model, which is essential for reducing position bias. 
Through aligning distance metrics across original and permuted lists, our model effectively identifies and mitigates bias, enhancing the robustness and efficacy of the ranking process.
The contributions of this paper are:
\begin{itemize}[noitemsep, topsep=0pt, partopsep=0pt]
    \item  We harmonize the goals of language generation and ranking tasks within a listwise framework using a novel soft lambda rank approach that incorporates explicit feedback, ensuring seamless integration of these objectives.
    \item We introduce a permutation-sensitive learning methodology that addresses position bias efficiently, without adding extra computational load during inference.
    \item We assess the performance of our model across four extensively used datasets, demonstrating its effectiveness.
\end{itemize}

\section{Related Works}
\subsection{Large Language Model for Recommendation}

Recent advancements in Large Language Models have showcased their formidable capabilities across a spectrum of tasks, drawing interest towards their potential in recommender systems~\cite{qiu2021u, bao2023tallrec, dai2023uncovering, zheng2024harnessing, DBLP:conf/aaai/WuQZZC24,DBLP:journals/corr/abs-2307-02157,
DBLP:conf/nips/ChenZCDLLD21, DBLP:journals/tois/ZhengWXSQZHWC23}. 
A comprehensive survey by~\citet{wu2023survey} listed the existing works on LLM-based Recommendations, particularly focusing on LLMs as agents that directly generate predictive outcomes. 
We delineated them into three paradigms, point-wise, pair-wise, and list-wise approaches.

The point-wise paradigm is characterized by the LLM processing each historical and candidate item pair individually.~\cite{liu2021jizhi, sachan2022improving, 
DBLP:conf/recsys/ZhengSSZ023,
zheng2024harnessing}
For example, \citet{bao2023tallrec} adapted the recommendation template to frame it as a yes-no question, requiring the LLM to evaluate each candidate sequentially. 
Another significant contribution is by \citet{li2023e4srec} and  \citet{yue2023llamarec}, who leveraged LLMs to recommend items through an adapter module that computes the probability of each item for recommendation.
In the pair-wise paradigm, the LLM determines the preferable option between two candidate items. \citet{qin2023large} introduced a pair-wise prompting strategy employing a sliding window technique to identify the recommended items.
Nonetheless,  the point-wise and pair-wise approaches are notably inefficient due to the necessity of repeatedly calling the LLM, escalating the time cost as the number of candidates increases~\cite{kang2023llms}.
In contrast, the listwise approach offers a more efficient solution by ranking the entire list of candidates in a single inference phase. 
Although some studies propose a listwise approach~\cite{Sun2023IsChatGPT, dai2023uncovering, yang2023palr, ma2023zero, drozdov2023parade, yue2023llamarec}, they often address the problem with supervised fine-tuning, while falling short in handling the rating magnitude with metric-oriented LLM-based recommendation.


\subsection{Learning to Rank}


Learning to Rank (LTR) constitutes a fundamental component in information retrieval systems, aimed at ordering entities by their relevance. This domain is categorized into three main methodologies according to the design of the loss function: pointwise, pairwise, and listwise approaches. Pointwise methods focus on predicting the absolute relevance of individual items, typically framed as classification or regression tasks~\cite{li2007mcrank, crammer2001pranking}. Pairwise strategies, in contrast, emphasize the relative importance between item pairs, to accurately determine the more relevant item in a pair~\cite{freund2003efficient, burges2005learning, chapelle2010efficient}. The listwise approaches extend this concept by considering the entire item list as the training unit, aiming to directly optimize the overall item ordering to align with ranking objectives~\cite{xu2007adarank, cao2007learning, taylor2008softrank, xia2008listwise, burges2010ranknet, DBLP:journals/tois/WangZZQCX24}.
In this paper, we present an innovative adaptation of the lambda loss function~~\cite{wang2018lambdaloss} tailored for natural language generation, leveraging the pairwise approach to enhance the coherence of generated texts. This adaptation underscores the potential of LTR methodologies to extend beyond traditional retrieval tasks.
\section{Problem Statement}


We define the sequential recommendation ranking problem as follows. Let $\mathcal{U}$ represent the set of users and $\mathcal{I}$ denote the set of items. For any given user $u \in \mathcal{U}$, their historical interactions with items are represented by $\mathcal{H}_u = \{h_1, h_2, \ldots, h_k\}$, where each $h_i \in \mathcal{I}$ signifies an item that user $u$ has previously interacted with. With this notation in place, the ranking problem is formalized as follows:

\begin{myDef}
For a user $u$, consider $\mathcal{C}_u = \{c_1, c_2, \ldots, c_m\}$ as the set of candidate items for recommendation, where each $c_i \in \mathcal{I}$ and $m \leq |\mathcal{I}|$. The goal is to devise a ranking function $F: \mathcal{H}_u \times \mathcal{C}_u \rightarrow S_m$ that accurately predicts the permutation $\tau \in S_m$ that best orders the items in $\mathcal{C}_u$. The set $S_m$ is the symmetric group of all $m$-element permutations, encapsulating every possible arrangement of the candidate items.
\end{myDef}



\section{Methodology}

In this section, we elucidate the constraints inherent in prevailing prompting paradigms when addressing list-wise recommendation tasks.  Our learning framework is developed with four distinct components: Template Design, Supervised Fine-Tuning, Soft Lambda Loss, and Permutation-Sensitive Learning.

\subsection{Template Design}
Before delving into the specifics of our learning module, we delineate the process of transforming the ranking task into a language generation problem. 
Drawing inspiration from Instruction Tuning \cite{alpaca}, we employ a natural language prompt template, denoted as $T_{\text{src}}(\mathcal{H}_u, \mathcal{C}_u)$, which transmutes the input user history $\mathcal{H}_u$ and context $\mathcal{C}_u$, inclusive of item attributes such as names, categories, and descriptions and their explicit rating from user, into a structured format. 
This transformation additionally aids in creating target text templates $T_{\text{tgt}}(\tau)$, representing the permutation that arranges candidate items according to user preferences.
The detailed template design and example are provided in Appendix~\ref{appendix:Template Design}.

\subsection{Supervised Fine-Tuning}
With the language generation problem that given $T_{\text{src}}(\mathcal{H}_u, \mathcal{C}_u)$ that aims to predict $T_{\text{tgt}}(\tau)$, we implement a supervised fine-tuning paradigm that leverages the Low-Rank Adaptation (LoRA) approach, as introduced by~\citet{hu2021lora}. 
The core idea behind LoRA is to adapt pre-trained models in a parameter-efficient manner, enabling effective fine-tuning on downstream tasks with minimal modifications to the original model parameters. 
The fine-tuning process is formulated by the following loss function:

\begin{equation}
    \mathcal{L}_{\text{sft}} = - \sum_{t=1}^{|y|} \log \left( P_{\theta}(y_t | x, y_{<t}) \right),
\end{equation}
where $\mathcal{L}_{\text{sft}}$ denotes the supervised fine-tuning loss, and $P_{\theta}(y_t | x, y_{<t})$ represents the conditional probability of predicting the token $y_t$ given the input tokens $x$ and the preceding tokens $y_{<t}$. In this context, $x$ and $y$ correspond to the tokenized representations of $T_{\text{src}}(\mathcal{H}_u,\mathcal{C}_u)$ and $T_{\text{tgt}}(\tau)$, respectively.
This supervised fine-tuning process utilizes target tokens that correspond to the correctly ranked list of candidate answers, which are subsequently adjusted to reflect user preferences. 


\subsection{Soft Lambda Loss (SLL) }
The widely adopted cross-entropy loss in language generation, derived from next-token prediction during supervised fine-tuning, faces a fundamental misalignment with the objectives of ranking. 
Such a discrepancy undermines the efficacy of cross-entropy loss when applied to the specific demands of ranking, leading to suboptimal performance in these contexts. To empower the Language Model with the capability to identify partial order relations, learning to rank (LTR) objectives serves as an effective supervised signal.
Unlike the existing LTR framework \cite{wang2018lambdaloss}, this is not straightforward to directly optimize on Normalized Discounted Cumulative Gain (NDCG) when dealing with language models that generate ranked token probabilities incrementally. 
Traditional ranking losses, such as Lambda loss~\cite{wang2018lambdaloss} or SoftRank~\cite{taylor2008softrank}, are not directly applicable. The Lambda loss, is defined as:
\begin{equation}
\begin{split}
\mathcal{L}_{\text{rank}} = \sum_{i=1}^{|\tau|} \sum_{\substack{j:\tau_j < \tau_i}} \delta_{i,j}|G_i - G_j| \cdot \\
\log_2 \left(1 + e^{-\sigma(s_i - s_j)}\right),
\label{eq:softlambda}
\end{split}
\end{equation}
where
\begin{equation}
   \delta_{ij} = \left| \frac{1}{D_{|i-j|}} - \frac{1}{D_{|i-j|+1}} \right|,
\end{equation}
with $G_i$ and $D_i$ following the definitions from NDCG, and $s_i$ representing the model-derived prediction score. 
In large language models, the ranking order is typically determined by using the $argmax$ function on the output probabilities of tokens, which is non-differentiable and thus unsuitable for the training process.

To overcome this, we propose a method that combines the soft-argmax function with Lambda loss to calculate the deviation of predicted probabilities from the ideal ranking order. We define a differentiable ranking score for the generative model by substituting the traditional $argmax$ function in $s_i$ with the soft-argmax, expressed as:
\begin{equation}
\label{eq:softargmax}
    s_i =\max_{j} \frac{e^{\gamma y_{j,i}}}{\sum_{k} e^{\gamma y_{j,k}}} \cdot j,
\end{equation}
where $y_{i,j}$ denotes the output probability of the language model for the $j$th position and token $i$, and $\gamma$ represents the scaled value that adjusts the distribution of softmax. By making the computation of $s_i$ differentiable with the soft-argmax method, we align the objectives of language generation with those of the ranking task. Overall, Soft Lambda Loss follows the Equation~\ref{eq:softlambda}, which is derived from \citet{wang2018lambdaloss}, by replacing $s_i$ with Equation~\ref{eq:softargmax} to get a differentiable objective.

\subsection{Permutation-Sensitive Loss (PSL)}

\begin{figure}[t]
    \centering
    \includegraphics[width=0.48\textwidth]{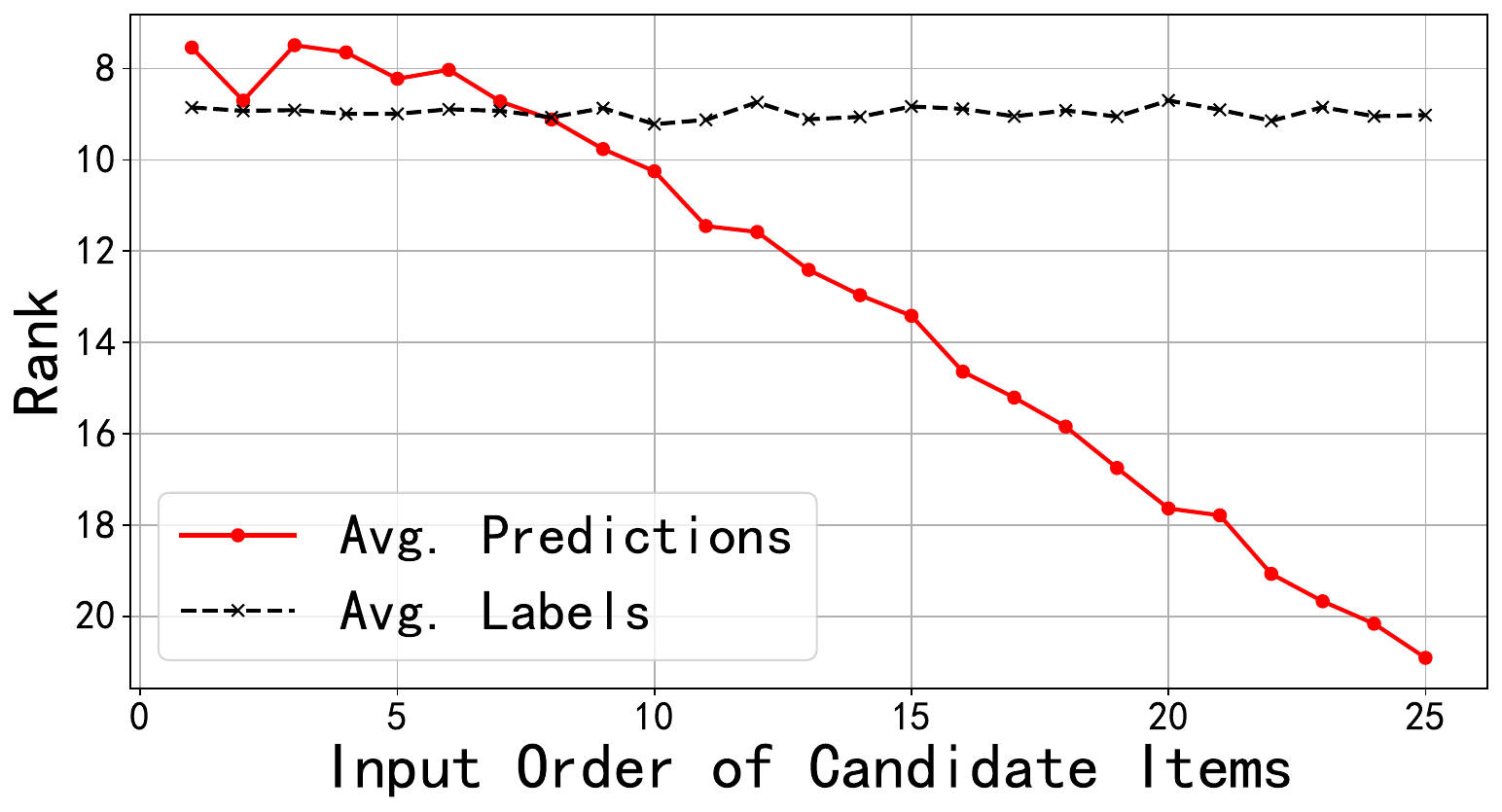}
    \caption{Demonstration of position bias. The figure shows how the placement of candidate items in the input sequence can significantly alter the ranking results produced by a Language Model.}
    \label{fig:positionbias}
\end{figure}

In list-wise recommendation tasks with Large Language Models, position bias emerges as a formidable challenge, with the order of the candidate input sequence notably swaying the ranking outcomes. 
As depicted in Figure~\ref{fig:positionbias}, language models exhibit a propensity to assign higher rankings to candidates positioned at the beginning of the list. This tendency highlights the significant influence of candidate positioning on model evaluations, underscoring the imperative of developing methodologies to counteract these biases.

It is worth noting that the observed phenomenon depends exclusively on natural language generation tasks with the sequence of input candidates. This contrasts with embedding-based recommendation systems, where the order of inputs does not influence outcomes by calculating the score of the user and item pair separately. The effect of permutation on the output is described by the inequation:
\begin{equation}
\label{eq:position}
F(T(\mathcal{H}_u,  \mathcal{C}_u)) \neq F(T(\mathcal{H}_u,  \mathcal{C}'_u)),
\end{equation}
where $F(\cdot)$ denotes the logits output by large language model, and $\mathcal{C}'_u = \{c_{\pi(0)}\,c_{\pi(1)}, \cdots, c_{\pi(m)}\}$ represents a permuted candidate list from the original candidate list $\mathcal{C}_u$, with $\pi(\cdot)$ as the random permutation function that rearranges the candidates. 
This equation highlights the dependency of the model on the sequence in which inputs are provided, distinguishing it from conventional recommendation approaches which are order invariant.

Although \citet{hou2023large} proposed the bootstrapping method, which shuffles the candidate items multiple times and takes average scores as the final ranking result, it is inefficient as it repetitively calls language models in the inference stage to get average ranking. 
To alleviate this issue without burdening the inference in the recommendation, we propose a permutation-sensitive loss that aims to minimize the output distribution distance between the original candidate list $\mathcal{C}_u $ and the random permutated candidate list $\mathcal{C}'_u $ within the fine-tuning stage. .
By adopting Kullback–Leibler divergence that minimizes the distance between two distributions, we empower the model with permutation invariant capability. The loss function could be formulated as:
\begin{equation}
\label{eq:kl}
\mathcal{L}_{\text{perm}} = \sum_{t} \text{KL} \left( P_{\theta}(y_t|x, y_{<t}) \| P_{\theta}(y'_{t}|x', y'_{<t}) \right),
\end{equation}
where $x$ and $x'$ are the prompt derived from $T(\mathcal{H}_u,\mathcal{C}_u)$ and $T(\mathcal{H}_u,\mathcal{C}'_u)$ respectively, and $y$ and $y'$ are the labels for the given prompts. The details of $\mathcal{C}'_u $, $y'_t$ and corresponding $P_{\theta}(y'_t|x',y'_{<t})$ are provided in Appendix~\ref{appendix:PSL}.

\subsection{Training Objective}

Overall, we provide the soft lambda loss $\mathcal{L}_{\text{rank}}$ with permutation-sensitive framework $\mathcal{L}_{\text{perm}}$ to address the issues mentioned above, which goes beyond the naive supervised fine-tuning.
The objective function is reformulated as: 
\begin{equation}
\label{eq:equation}
\mathcal{L} =  \mathcal{L}_{\text{sft}} + \alpha\mathcal{L}_{\text{rank}} + \beta\mathcal{L}_{\text{perm}},
\end{equation}
where $\alpha, \beta$ are hyperparameters that adjust the importance of each loss.
\section{Experiment}
In our study, we conducted a comprehensive evaluation of our model across two real-world datasets. 
This was complemented by an ablation study, robustness tests, and efficiency evaluations. Our experiment was directed by the following pivotal research questions:
\begin{itemize}[noitemsep, topsep=0pt, partopsep=0pt]
    \item \textbf{(RQ1)} Does the proposed framework surpass existing baselines in both embedding-based and LLM-based recommendation models?
    \item \textbf{(RQ2)} What extent does supervised fine-tuning on recommendation-specific corpus enhance Large Language Model performance?
    \item \textbf{(RQ3)} How crucial is the involvement of our proposed module for metrics improvement?
    \item \textbf{(RQ4)} How does permutation-sensitive learning compare to bootstrapping methods in terms of performance and efficiency?
    \item \textbf{(RQ5)} How does the ALRO framework improve performance across different parameter sizes of the backbone language model compared to traditional supervised fine-tuning?
\end{itemize}

Through these explorations, we aim to elucidate the contributions of domain-specific fine-tuning with our novel modules to the advancements in LLM-based recommendation systems.

\subsection{Dataset}

We selected four widely adopted open-source datasets to evaluate the effectiveness of our framework: \textit{Movie} (MovieLens-1M\footnote{\url{https://grouplens.org/datasets/movielens/1m/}}), \textit{Music} (the "CDs \& Vinyl" subset), \textit{Books} (the "Books" subset), and \textit{Games} (the "Toys and Games" subset) from the Amazon product reviews dataset. The Amazon product reviews datasets encompass reviews from 1996 to 2023~\cite{hou2024bridging} with 5-core. Detailed information about these datasets is presented in Table~\ref{tab:dataset}.
To evaluate the model's capability of ranking explicit feedback, we sampled the most recent 25 user-interacted items as candidates $\mathcal{C}_u$, each with a rating $r_{c_i} \in [1\mathrel{{.}\,{.}}\nobreak5]$. The output permutation $\tau$ is sorted from the candidate ratings $r_{c_i}$ provided by the user. The length of historical sequence $|\mathcal{H}_u|$ is set to 20. Followed by~\citet{kang2018self}, we split the user interaction sequence into three-part, 1) the most recent 25 actions for testing 2) the most recent 25 to 50 actions for validation 3) all remaining actions for training.

\begin{table}[t]
    \caption{Statistics of datasets.}
    \small
    \centering
    \begin{tabular}{c|cccc}
    \hline 
    Dataset & Movie & Books & Games & Music  \\
    \hline 
    Users & 6040 & 54440 & 12182 & 9612  \\
    Items & 3952 & 446987 & 114601 & 83937  \\
    Actions & 1.0M & 9.27M  & 1.53M & 1.58M \\
    Density (\%) & 4.19 & 0.0381 & 0.11 & 0.197  \\
    Tokens/Item & 20.76 & 45.53 & 56.92 & 24.10  \\
    \hline
    \end{tabular}
    \vspace{-3pt}
    \label{tab:dataset}
\end{table}

\subsection{Baselines and Evaluation Metrices}

To evaluate the effectiveness of our framework, we select several state-of-the-art baselines, which could be categorized into Non-Sequential Recommendation, Sequential Recommendation, Ranking Methods, and Large Language Model-based Recommendation. We introduce the BERT-based model as the backbone to extract the textual information of items in both Non-Sequential Recommendation and Sequential Recommendation.
\begin{itemize}[noitemsep, topsep=0pt, partopsep=0pt]
    \item \textbf{Non-Sequential Recommendation:} \textbf{NCF}~\cite{he2017neural} adopts a neural network with collaborative filtering for recommendations. \textbf{DIN}~\cite{kang2018self} involves user interest modeling based on user behavior with an attention mechanism.
    \item \textbf{Sequential Recommendation:} \textbf{GRU4Rec}~\cite{hidasi2015session} is a session-based recommendation system utilizing a GRU-based recurrent network. \textbf{SASRec}~\cite{hidasi2015session} employs a self-attention network with positional embeddings to capture the user's sequential behavior information. \textbf{CORE}~\cite{hou2022core} uses a representation-consistent framework to unify the session and item representation spaces. \textbf{NARM}~\cite{li2017neural} decomposes user behavior into global and local forms using attention networks for sequential recommendation.
    \item \textbf{Ranking Methods:} \textbf{Seq2Slate}~\cite{bello2018seq2slate} adopts RNN modules with a pointer network that maps candidate items to ranking positions in an end-to-end manner. \textbf{PRM}~\cite{pei2019personalized} utilizes a transformer-based network to re-rank lists by assigning scores to each candidate in a list-wise form.
    \item \textbf{Large Language Model-based Recommendation:} For \textbf{Zero-shot LLM} and \textbf{Few-shot LLM}, we follow list-wise setting in \citet{hou2024large}, which provides instructions and examples. \textbf{TallRec} \cite{bao2023tallrec} fine-tunes LLMs with instruction tuning for point-wise recommendation. \textbf{ES4Rec} \cite{li2023e4srec} introduces pre-trained item embeddings as prompts with an adapter to fine-tune the LLM.  \textbf{LlamaRec} \cite{yue2023llamarec} employs a two-stage re-ranking framework for recommendation. We adopted the LLM re-ranking module in LlamaRec for ranking the candidates. We use Llama2-7b as the base model for all LLM-based baselines. It is worth noting that \textbf{ES4Rec} and \textbf{TallRec} require negative sampling data to maintain the performance of learning user embeddings, which imposes an additional burden on LLM training.
\end{itemize}

To assess the performance of various models in ranking tasks for explicit feedback (value from 1 to 5), we employ Normalized Discounted Cumulative Gain (NDCG) at different cutoffs as our evaluation metric, specifically NDCG@k with $k$ values of 3, 5, 10, and 25.

\begin{table*}[tb]  
\centering  
\small
\caption{Performance Comparison. Optimal outcomes across all models are emphasized in bold, while second-best performances are distinguished by underlining. Evaluation metrics include NDCG at ranks 3, 10, and 25.}  
\label{tab:main_result}  
  \begin{center}  
  \resizebox{\textwidth}{!}{
      \begin{tabular}{l|ccc|ccc|ccc|ccc}  
        \toprule
        \textbf{Dataset}&\multicolumn{3}{c|}{\textbf{Movie}}&\multicolumn{3}{c|}{\textbf{Books}}&\multicolumn{3}{c|}{\textbf{Games}}&\multicolumn{3}{c}{\textbf{Music}}\\\midrule
        
        \textbf{NDCG}&@3&@10&@25&@3&@10&@25&@3&@10&@25&@3&@10&@25 \\
        \midrule
        \textbf{NCF}& 0.5804&0.6452&0.8336&0.7482&0.7692&0.9040&0.8245&0.8346&0.9356&0.7733&0.7918&0.9140\\
        \textbf{DIN}& 0.6067&	0.6674&	0.8437&0.7495&0.7711&0.9048&0.8240&0.8337&0.9351&0.7804&0.7946&0.9153\\
        \midrule
        \textbf{GRU4Rec}& 0.5545	&0.6268&	0.8241&0.7461&0.7679&0.9034&0.8258&0.8344&0.9355&0.7705&0.7858&0.9117\\ 
        \textbf{DIEN}& 0.5890&	0.6580&	0.8385&0.7508&0.7716&0.9051&0.8331&0.8376&0.9372&0.7728&0.7919&0.9133\\
        \textbf{SASRec}&0.6436&	\underline{0.6891}&	0.8427&0.7796&0.7961&0.9153&\underline{0.8501}&0.8533&\underline{0.9432}&0.7979&0.8135&0.9229
\\
        \textbf{COREave}&0.6236&	0.6455&	0.8299&0.7740&0.7899&0.9128&0.8498&0.8503&0.9409&0.7876&0.7957&0.9146\\
        \textbf{NARM}&0.5281&	0.6059&	0.8145&0.7285&0.7555&0.8979&0.8282&0.8373&0.9366&0.7679&0.7863&0.9118\\
        \midrule
        \textbf{Seq2Slate}& 0.5320&  0.6034& 0.8178 & 0.7850&0.7961&\underline{0.9160}&0.8292&0.8345&0.9357&0.7850&0.7961&0.9160\\
        \textbf{PRM}& 0.6088& 0.6496& 0.8384&0.7668&0.7858&0.9116&0.8235&0.8315&0.9344&0.7668&0.7858&0.9116\\
        \midrule
        \textbf{Zero-shot}& 0.5118&	0.5954&	0.8089&0.7322&0.7576&0.8989&0.8190&0.8317&0.9339&0.7639&0.7841&0.9106\\
        \textbf{Few-shot}& 0.5149&	0.5958&	0.8097&0.7337&0.7595&0.8995&0.8288&0.8358&0.9362&0.7746&0.7874&0.9128\\
        \textbf{TALLRec}&\underline{0.6512}&	0.6835&	\underline{0.8494}&\underline{0.7895}&\textbf{0.8153}&0.9141&0.8492&0.8469&0.9397&\underline{0.8221}&\underline{0.8269}&\textbf{0.9295}\\
        \textbf{E4SRec}& 0.5697&	0.6210&	0.8373&0.7620&0.7878&0.9089&0.8477&\underline{0.8545}&0.9354&0.7795&0.7951&0.9203\\
        \textbf{LlamaRec}& 0.5360&	0.6164&	0.8193&0.7439&0.7687&0.9092&0.8402&0.8458&0.9366&0.7831&0.7945&0.9154\\
        
        \midrule
        \textbf{ALRO}&\textbf{0.6584}&	\textbf{0.6925}&	\textbf{0.8590}&\textbf{0.7903}&\underline{0.7981}&\textbf{0.9190}&\textbf{0.8555}& \textbf{0.8582}& \textbf{0.9472} & \textbf{0.8310} & \textbf{0.8411} & \underline{0.9283}\\
        \bottomrule
      \end{tabular}}  
  \end{center}  
         
\end{table*}

\subsection{Implementation Details}
Our experiments were conducted on a cluster of 12 Linux servers, each equipped with 8 A800 GPUs. For the backbone model, we utilized the Llama2-7b~\footnote{https://huggingface.co/meta-llama/Llama-2-7b-hf} with BF16 precision, available on Huggingface. The supervised fine-tuning step was implemented using the PyTorch framework and peft library, applying the LoRA technique with a rank setting of 16. We used the AdamW~\cite{loshchilov2017decoupled} optimizer with a learning rate of 5e-5 and batch size as 128 for SFT, complemented by 2 gradient accumulation steps with a total of 10 training epochs. We utilized DeepSpeed with ZeRO stage 2 to facilitate distributed training. To optimize our loss function, we performed hyperparameter tuning using a grid search across the values of $\gamma$ and $\beta$ within the set  {0.001,0.01,0.1,1,2,5}. We fixed $\alpha$ at 1, $\beta$ at 2, and $\gamma$ at 2 during this process. 

\subsection{Overall Performance (RQ1)}
To validate the performance of our proposed framework, ALRO, we executed comparative analyses against established baseline methods, with the results presented in Table~\ref{tab:main_result}. The following observations were made:

\begin{itemize}[noitemsep, topsep=0pt, partopsep=0pt]
    \item ALRO consistently outperformed the baselines across various metrics and datasets, unequivocally demonstrating its superiority in ranking tasks within recommender systems.
    \item Large Language Models (LLMs) without fine-tuning fell short against traditional methods, highlighting the crucial role of supervised fine-tuning for LLMs in recommendation contexts.
    \item TALLRec achieves comparable performance but faces efficiency challenges.
\end{itemize}

These insights confirm the significance of our ALRO framework in enhancing the efficacy of ranking in recommendation systems and underscore the necessity for appropriate fine-tuning of LLMs to fully leverage their potential recommendation.

\subsection{Effect of Supervised Fine-Tuning (RQ2)}
\begin{table}[t]  
    \centering  
    \caption{Comparison of Zero-shot, Few-shot and Supervised Fine-Tuning with Llama2-7b backbone.}  
    \label{tab:supervised_fine_tuning}  
      \begin{center}  
          \begin{tabular}{l|ccc}  
            \toprule
            \textbf{Dataset}&\multicolumn{3}{c}{\textbf{Movie}}\\ 
            \midrule
            \textbf{NDCG}&@3&@10&@25\\
            \midrule
            \textbf{Zero-shot}& 0.5118&	0.5954&	0.8089\\
            \textbf{Few-shot}& 0.5149&	0.5958&	0.8097\\
            \textbf{SFT} &\textbf{0.5712}&\textbf{0.6413}&\textbf{0.8258}\\ 
            \bottomrule
          \end{tabular}  
      \end{center}  
\end{table}
Prompting techniques have showcased the profound ability of language models to interpret and execute tasks with remarkable precision.~\cite{liu2023pre} However, the efficacy of these techniques is challenged when applied to specialized domains such as recommendation systems, particularly due to the potential misalignment between the pre-training corpus and the intricate requirements of ranking tasks.
As depicted in Table~\ref{tab:supervised_fine_tuning}, this discrepancy is notably pronounced in medium-sized language models like Llama-7b, where simple prompting may not suffice to activate the model's ranking capabilities effectively.

To address this gap, our study delves into the impact of supervised fine-tuning on the performance of language models in recommendation-related tasks. 
Through a comparative analysis encompassing zero-shot, few-shot, and supervised fine-tuning approaches, we unveil a substantial improvement in model performance by supervised fine-tuning, with metrics enhancing by over 10\%. 
This improvement is attributed to the fine-tuning process, which effectively adjusts the model's outputs to better align with specific task requirements. This approach overcomes the shortcomings of conventional prompting techniques that often yield non-parsable outputs, thereby enhancing the model's ability to rank information more accurately.
\subsection{Ablation Study (RQ3)}
\begin{figure}
\centering
\subfloat[Movie dataset.]{
    \includegraphics[width=0.45\linewidth]{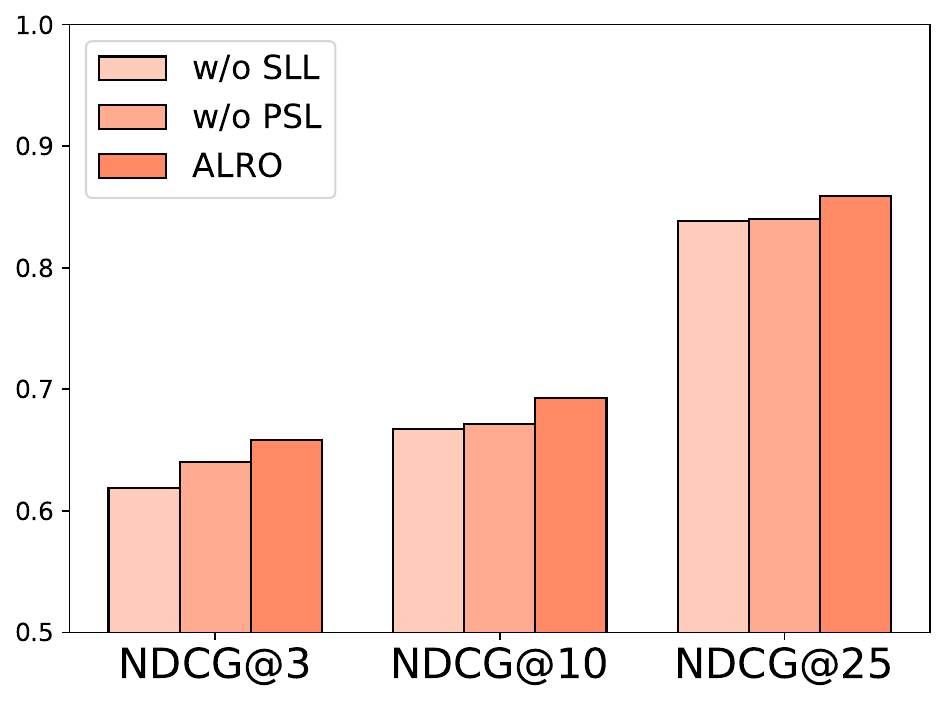}
    \label{fig:ablation_study_Movie}
     \vspace{-2mm}
}
\hfill
\subfloat[Music dataset.]{
    \includegraphics[width=0.45\linewidth]{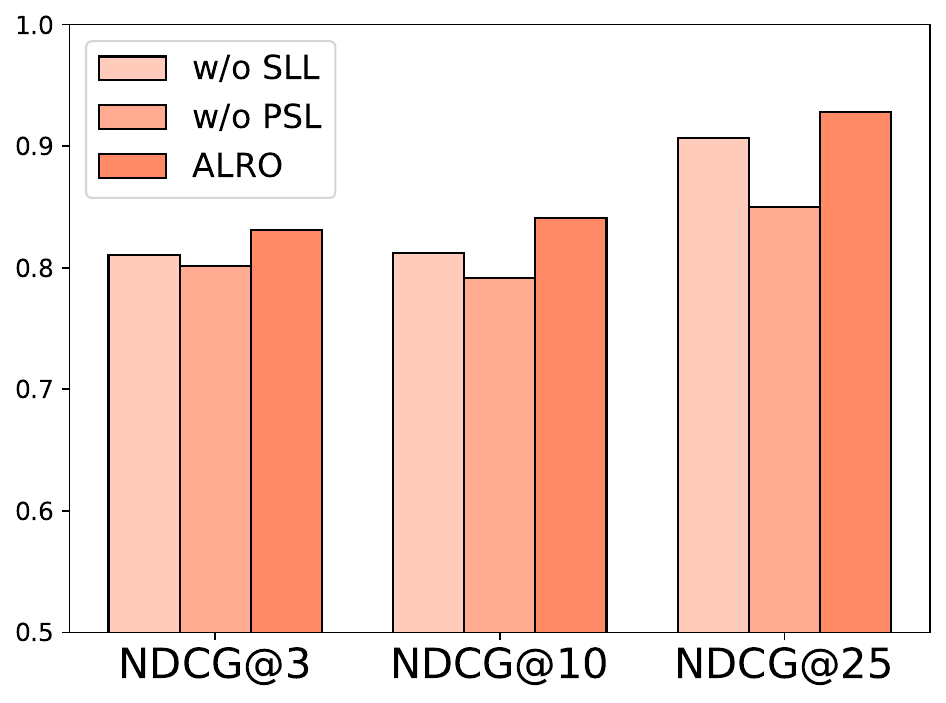}
    \label{fig:ablation_study_Music}
    \vspace{-2mm}
}
\caption{Ablation study on multiple datasets.}
\label{fig:ablation_study}
    
\end{figure}

In our research, we conducted an ablation study to distinguish the contributions of distinct components within our proposed framework, systematically omitting each module for comparative analysis against the complete model. This involved evaluating two key variants: Exclusion of soft lambda loss (w/o SLL) and Exclusion of permutation-sensitive learning (w/o PSL).
Figure~\ref{fig:ablation_study} shows that both components significantly enhance the system's candidate ranking ability.
The reduction in NDCG is attributed to the exclusion of the soft lambda loss, highlighting the importance of objective alignment in enhancing language models as recommender systems. Additionally, the performance drop from removing Permutation-Sensitive Learning underscores the impact of position bias on ranking performance.

\subsection{Comparison of Bootstrapping and Permutation-Sensitive Learning (RQ4)}

\begin{table}[t]  
    \centering  
    \caption{Comparative analysis of bootstrapping and permutation-sensitive learning. `p@i' denotes the number of permutations applied in bootstrapping. The original permutation represented by p@1 is consistent with the prompt in ALRO. TPD represents the average inference time per data sample, measured in seconds.}  
    \label{tab:PSL}  
      \begin{center}  
          \begin{tabular}{l|ccc|c}  
            \toprule
            \textbf{Dataset}&\multicolumn{4}{c}{\textbf{Movie}}\\ 
            \midrule
            \textbf{NDCG}&@3&@10&@25& TPD \\
            \midrule
            \textbf{p@1}& 0,6004& 0.6203& 0.8156& 0.2546 \\
            \textbf{p@3}& 0.6217& 0.6842& 0.8554& 0.7654\\
            \textbf{p@5}& 0.6472& \textbf{0.7032}& \textbf{0.8646}& 1.3756\\
            \textbf{ALRO} &\textbf{0.6584}&0.6925&0.8590& 0.2546\\ 
            \bottomrule
          \end{tabular}  
      \end{center}  
         
\end{table}

Our research introduces a permutation-sensitive learning approach designed to address position bias, which affects the outcomes based on the order of candidate lists. While the bootstrapping method~\cite{hou2023large} , offers a solution to this bias, it significantly increases inference time. We evaluated the effectiveness of permutation-sensitive learning compared to bootstrapping, aiming to reduce position bias without burdening the inference stage. Our comparisons included the original model without modifications, and bootstrapping with permutations executed 3 and 5 times. As demonstrated in Table~\ref{tab:PSL}, our method achieves comparable outcomes to bootstrapping while reducing inference times by approximately 5-fold. This indicates that our approach effectively mitigates the inference time issue through well-designed learning objectives.

\subsection{Effect of Model parameter size (RQ5)}

\begin{figure}[t]
    \centering
    \includegraphics[width=0.48\textwidth]{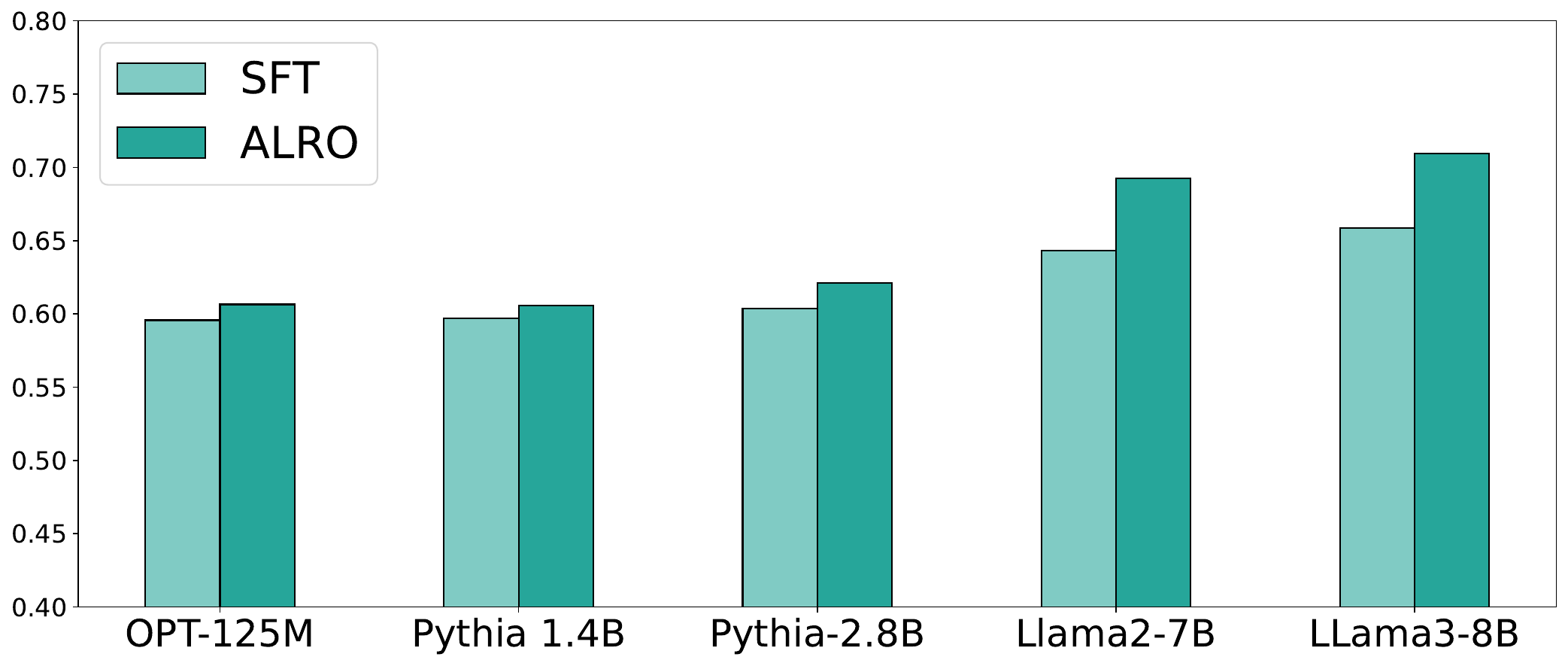}
    \caption{Enhancements achieved by ALRO across various model sizes on Movie dataset, measured using NDCG@10 metric.}
    \label{fig:parameter_size}
\end{figure}

In this section of our research paper, we delve into the adaptability and efficacy of our learning framework across several LLM-based recommender systems, spanning various model sizes. 
Specifically, we selected four distinct models for our analysis: OPT-125M, Pythia 1.4B, Pythia-2.7B, Llama2-7B, Llama3-8B. 
By applying our framework to these models, we aim to showcase the consistent and significant performance enhancements it offers compared to traditional supervised fine-tuning approaches. 
As depicted in Figure~\ref{fig:parameter_size}, there is a clear correlation between model parameter size and performance, which serves to emphasize the capacity of our learning framework to augment the effectiveness of recommender systems across a spectrum of language model sizes. 
Notably, the enhancements provided by our framework are more significant in larger models than in smaller ones, this may be attributed to the innate reasoning capability of language models. Overall, the experiment highlights the versatility and broad applicability 
of our framework in improving system performance.
\section{Conclusion}

In this research, we tackled the intricacies of employing large language models as ranking agents in recommender systems with explicit feedback, focusing on refining list-wise ranking methods to manage the order relation. 
We proposed a cutting-edge framework that integrates soft lambda loss and permutation-sensitive learning.
The integration of soft lambda loss is important as it bridges the objective between LLM's natural language generation and the specific demands of ranking tasks. It enhances the performance of ranking by optimizing the order relation within the magnitude of ratings.
Furthermore, permutation-sensitive learning approaches effectively address the issue of position bias, providing an improvement over traditional bootstrapping methods without imposing additional computational demands during inference. 
Our comprehensive evaluation across various datasets confirms the success of our method, advancing LLMs as recommendation agents.
\section{Limitation}


While our framework adeptly aligns the objectives of ranking and language generation, it falls short in fully harnessing the explainability potential inherent in language models. The supervised fine-tuning process, augmented by joint loss optimization, effectively enhances the model's performance in list-wise ranking tasks, particularly in recommendation systems. However, this process inadvertently undermines the model's proficiency in tasks beyond recommendation, limiting its versatility. Furthermore, although our method demonstrates efficacy in ranking a set of 25 items, scalability becomes a concern as the number of candidates increases significantly. This limitation arises due to constraints such as context limits or the propensity for forgetting in Large Language Models, compromising the model's ability to maintain performance consistency across varying candidate sizes.  Typically, when dealing with large candidate sets, methods such as sliding windows~\cite{Sun2023IsChatGPT} or retrieve-and-rank two-stage approaches~\cite{yue2023llamarec} are employed to address scalability issues.

\section{Acknowledgement}
This work was supported by the National Natural Science Foundation of China (Grant No.62102110, No.92370204), National Key R\&D Program of China (Grant No.2023YFF0725004), Guangzhou-HKUST(GZ) Joint Funding Program (Grant No.2023A03J0008), Education Bureau of Guangzhou Municipality.
\bibliography{main}

\appendix
\clearpage
\begin{appendices}
\section{Appendix}
\subsection{Template Design}
\label{appendix:Template Design}

We followed the template design from existing works~\cite{bao2023tallrec,yue2023llamarec} and refined the prompt to rank the items in a list-wise manner and alleviate position bias, as shown in Table~\ref{tab:Instruction Template}.
Specifically, the ranking results are sorted based on the rating \( r_{c_i} \in [1 \ldots 5] \). For candidates with equal ratings, we further sort them alphabetically. 
It is worth noting that while the equal rating results affect the supervised fine-tuning loss, they do not impact the soft lambda loss suggested in our framework, as the cumulative gain assigned in DCG for items with the same rating remains consistent.

\subsection{Permutation Sensitive Loss}
\label{appendix:PSL}
We generate the candidate list \(\mathcal{C}'_u = \{c_{\pi(0)}, c_{\pi(1)}, \cdots, c_{\pi(m)}\}\), which represents a permuted version of the original candidate list \(\mathcal{C}_u\), where \(\pi(\cdot)\) is a random permutation function. When the order of the candidate list is permuted, the corresponding target answer \(T_{\text{tgt}}(\tau')\) also noted as $y'_t$ is adjusted to match the new order. For example, referring to Table~\ref{tab:Instruction Template}, if we permute the candidates "Starman" and "Jumanji," the corresponding ranking result will change from "B A C ..." to "A B C ...". This permutation ensures that the model learns to rank based on the content rather than the position of the items in the list.



Regarding the probability distribution \(P_{\theta}(y'_t|x', y'_{<t})\), our objective is to minimize the distance of the output distribution after permutation. Let \(k_{\text{id}}\) represent the set of all token IDs and \(k_{\alpha}\) represent the set of alphabetic tokens. When the targeted alphabetic token ID \(k_{\alpha}\) changes according to the permutation function \(\pi(\cdot)\), we apply the same permutation function to adjust the token categories in the target distribution.
Mathematically, we aim to minimize the following loss:

\begin{equation}
\mathcal{L}_{\text{perm}} = \sum_{t} \text{KL} \left( P_{\theta}(y_t|x, y_{<t}) \| P_{\theta}(y'_{t}|x', y'_{<t}) \right),
\end{equation}
where \(\text{KL}(\cdot \| \cdot)\) denotes the Kullback-Leibler divergence, measuring the difference between the original output distribution and the permuted output distribution.

After applying the permutation \(\pi(\cdot)\) on the candidate, the output tokens ID \(k'_{\alpha}\) are changed. The permuted token ID set is \(k'_{\alpha} = \pi(k_{\alpha})\). Consequently, the target distribution must be adjusted to reflect the new order:

\begin{multline}
P_{\theta}(y'_{t, k'_{\alpha}}|x', y'_{<t, k'_{\alpha}}) \\ 
= P_{\theta}(y'_{t,\pi^{-1}(k_{\alpha})}|x', y'_{<t, \pi^{-1}(k_{\alpha})}).
\end{multline}
This means that the output distribution should accurately reflect the new order imposed by the permutation. The objective is to ensure that the output distribution of the permuted prompt \(P_{\theta}(y'_{t}|x', y'_{<t})\) closely matches the original distribution \(P_{\theta}(y_t|x, y_{<t})\), thereby maintaining the integrity of the ranking despite the permutation.

\begin{table}[t]
    \centering
    \small
    \caption{Instruction Template and Example}
    \label{tab:Instruction Template}
    \begin{tabular}{p{0.95\linewidth}}
        \toprule
        \textbf{Prompt Template} \\ 
        \midrule
        \#\#\# \textbf{Instruction:} \\
        Given the user’s interaction history, which reveals their items preferences, 
        generate a preference-based ranking of the provided candidate items. 
        Your task is to rank a list of new candidate movies. \\
        Your ranking should include all the candidate movies provided, and it should be based solely on the user's preferences, without regard 
        to the initial order of the candidates. \\
        \#\#\# \textbf{Input:} \\
        \textbf{[User Interaction History]:} \\
        <User Interaction History> \\
        \textbf{[Candidate Items]:} \\
        <Candidate Items> \\
        \#\#\# \textbf{Response:} \\
        Given the historical interaction, the ranking result is: \\
        <Ranking Result> \\
        \midrule
        \textbf{Example} \\
        \midrule
        \#\#\# \textbf{Instruction:} \\
        Given the user’s interaction history, which reveals their items preferences, 
        generate a preference-based ranking of the provided candidate items. 
        Your task is to rank a list of new candidate movies. \\
        Your ranking should include all the candidate movies provided, and it should be based solely on the user's preferences, without regard 
        to the initial order of the candidates. \\
        \#\#\# \textbf{Input:} \\
        \textbf{[User Interaction History]:} \\
        title: Independence Day genres: Action|SciFi|War rating: 3 \\
        title: Close Encounters of the Third Kind (1977) genres: Drama|Sci-Fi rating: 4 \ldots \\
        \textbf{[Candidate Items]:} \\
        (A) title: Starman genres: Adventure|Drama|Romance \\
        (B) title: Jumanji (1995) genres: Adventure|Children's |Fantasy \ldots \\
        \#\#\# \textbf{Response:} \\
        Given the historical interaction, the ranking result is: \\
        B A C \ldots \\
        \bottomrule
    \end{tabular}
\end{table}
\end{appendices}
\end{document}